\documentstyle[12pt]{article}
\begin{document}
\tolerance=5000
\def\be{\begin{equation}}
\def\ee{\end{equation}}
\def\bea{\begin{eqnarray}}
\def\eea{\end{eqnarray}}
\def\nn{\nonumber \\}
\def\beaa{\begin{eqnarray*}}
\def\eeaa{\end{eqnarray*}}
\def\cF{{\cal F}}
\def\det{{\rm det\,}}
\def\Tr{{\rm Tr\,}}
\def\e{{\rm e}}
\def\etal{{\it et al.}}
\def\erp2{{\rm e}^{2\rho}}
\def\erm2{{\rm e}^{-2\rho}}
\def\er4{{\rm e}^{4\rho}}
\def\etal{{\it et al.}}

\ 

\vskip -2cm

\ \hfill
\begin{minipage}{3.5cm}
October 1998 \\
\end{minipage}

\vfill

\begin{center}
{\Large\bf Weyl anomaly from Weyl gravity}

\vfill

{\sc Shin'ichi NOJIRI}\footnote{\scriptsize 
e-mail: nojiri@cc.nda.ac.jp, snojiri@yukawa.kyoto-u.ac.jp} and
{\sc Sergei D. ODINTSOV$^{\spadesuit}$}\footnote{\scriptsize 
e-mail: odintsov@mail.tomsknet.ru, odintsov@itp.uni-leipzig.de}

\vfill

{\sl Department of Mathematics and Physics \\
National Defence Academy, 
Hashirimizu Yokosuka 239, JAPAN}

\ 

{\sl $\spadesuit$ 
Tomsk Pedagogical University, 634041 Tomsk, RUSSIA \\
}

\ 

\vfill

{\bf abstract}

\end{center}

We calculate the conformal anomaly from 5d Weyl gravity (with 
broken conformal symmetry) which 
is conjectured to be supergravity dual to ${\cal N}=2$ superconformal 
field theory via AdS/CFT correspondence. Its comparison 
with ${\cal N}=2$ SCFT conformal anomaly (UV calculation) suggests 
that such duality may exist subject to presence of sub-leading 
$1/N$ corrections to cosmological and gravitational constants. 

\newpage

For years, Weyl gravity attracts the attention as 
an alternative theory of gravity. There have been suggested 
mechanisms how to get usual general relativity (low energy 
behavior) from it, for example, via breaking of Weyl (or conformal) 
invariance.

Even if Weyl gravity itself is not realistic theory it can be 
important as an essential element of other models. An example of 
that sort is provided by string theory. According to refs.\cite{Marc} 
there exists a scheme where the string-frame metric and dilaton 
dependent terms in low energy string action are given by 
\be
\label{F1}
\int d^{10}x \sqrt{G} \e^{-2\phi}\left[ R + 4 \left(\nabla_\mu 
\phi\right)^2 + c_1 W(R) \right]\ .
\ee
Here $c_1\sim {\alpha'}^3$ and $W(R)\sim C_{\mu\nu\alpha\beta}^4$.
This is caused by the field redefinition ambiguity \cite{GWT} which 
allows to change the coefficients of terms with Ricci tensor. In 
ref.\cite{Ts1} above action has been used to calculate strong 
coupling limit free energy for ${\cal N}=4$ super Yang-Mills 
theory via AdS/CFT correspondence (for an introduction, see \cite{AGM}).

The usual form of Weyl gravity is given by the 
square of the Weyl tensor 
$C_{\mu\nu\rho\sigma}C^{\mu\nu\rho\sigma}$
\be
\label{W1}
S=-\tilde c \int d^5 x \sqrt{- g}
C_{\mu\nu\rho\sigma}C^{\mu\nu\rho\sigma} \ .
\ee
In five dimensions, 
\be
\label{W2}
C_{\mu\nu\rho\sigma}C^{\mu\nu\rho\sigma}
= {1 \over 6} R^2 - {4 \over 3} R_{\mu\nu} R^{\mu\nu}
+ R_{\mu\nu\rho\sigma} R^{\mu\nu\rho\sigma}\ .
\ee
Since Weyl tensor is, of course, invariant under the Weyl 
transformation:
\be
\label{W3}
g_{\mu\nu}\rightarrow \e^{2\sigma}g_{\mu\nu}\ ,
\ee
the action (\ref{W1}) is Weyl invariant. 
The invariance is broken if we include the Einstein and 
cosmological terms:
\be
\label{W4}
S=-\int d^5 x \sqrt{- G}\left\{ {1 \over \kappa^2} R - \Lambda 
+  \tilde c C_{\mu\nu\rho\sigma}C^{\mu\nu\rho\sigma} \right\} \ .
\ee

The string theory dual to ${\cal N}=2$ superconformal field 
theory is presumably IIB string on ${\rm AdS}_5\times
X_5$ \cite{AFM} where $X_5=S^5/Z_2$. (The ${\cal N}=2$ $Sp(N)$ 
theory arises as the low-energy theory on the world volume on 
$N$ D3-branes sitting inside 8 D7-branes at an O7-brane). 

Then 
${1 \over \kappa^2}$ and $\Lambda$ are given by 
\be
\label{prmtr}
{1 \over \kappa^2}={N^2 \over 4\pi^2}\ ,\quad
\Lambda= -{12N^2 \over 4\pi^2}\ .
\ee
The Riemann curvature squared term in the above bulk action 
may be deduced from heterotic string via heterotic-type I duality 
\cite{Ts2}. Using field redefinition ambiguity \cite{GWT} 
one can suppose that there exists the scheme where 
$R_{\mu\nu\alpha\beta}^2$ may be modified to 
$C_{\mu\nu\alpha\beta}^2$ in the same way as in ref.\cite{Marc}. 
Then, the action (\ref{W4}) is presumably the bulk action dual 
to ${\cal N}=2$ SCFT. 

Our purpose here will be to check this conjecture, comparing the 
conformal anomaly in $Sp(N)$ ${\cal N}=2$ SCFT (UV calculation) 
and the conformal anomaly derived from the action (\ref{W4}) 
via AdS/CFT correspondence (SG side) \cite{Witten}. 

  Within AdS/CFT correspondence, the Weyl anomaly in 
4 dimensions has been calculated from the 5 dimensional 
higher derivative action in \cite{NO1}. 
In the conjecture of AdS/CFT correspondence, the partition 
function in $d$-dimensional conformal field theory is given 
in terms of the classical action in $d+1$-dimensional 
gravity theory:
\be
\label{W5}
Z_d(\phi_0)=\e^{-S_{\rm AdS}\left(\phi^{\rm classical}
(\phi_0)\right)}\ .
\ee
Here $\phi_0$ is the value of the field $\phi$ on the 
boundary and $\phi^{\rm classical}(\phi_0)$ is a field 
on bulk background, which is AdS, given by solving the 
equations of motion with the boundary value $\phi_0$ 
on $M^d$.$S_{\rm AdS}\left(\phi^{\rm classical}(\phi_0)
\right)$ is the classical gravity action on AdS. 
When we substitute the classical solution into the action, 
the action, in general, contains infrared divergences 
coming from the infinite volume of AdS.
Then we need to regularize the infrared divergence. 
It is known that as a result of the regularization and 
the renormalization there often appear anomalies.  
As fluctuations around the anti de Sitter space, 
we assume the metric has the following form:
\bea
\label{W6}
ds^2&\equiv&\hat G_{\mu\nu}dx^\mu dx^\nu 
= {l^2 \over 4}\rho^{-2}d\rho d\rho + \sum_{i=1}^d
\hat g_{ij}dx^i dx^j \nn
\hat g_{ij}&=&\rho^{-1}g_{ij}\ .
\eea
We should note that there is a redundancy in the expression 
of (\ref{W6}). In fact, if we reparametrize the metric :
\be
\label{W7}
\delta\rho= \delta\sigma\rho\ ,\ \ 
\delta g_{ij}= \delta\sigma g_{ij}\ .
\ee
by a constant parameter $\delta\sigma$, 
the expression (\ref{W6}) is invariant. The transformation 
(\ref{W7}) is nothing but the scale transformation on $M_d$.
We expand the metric $g_{ij}$ as a power 
series with respect to $\rho$,
\be
\label{vii}
g_{ij}=g_{(0)ij}+\rho g_{(1)ij}+\rho^2 g_{(2)ij}+\cdots \ .
\ee
We regard $g_{(0)ij}$ in (\ref{vii}) as independent field on 
$M_d$. We can solve $g_{(l)ij}$ ($l=1,2,\cdots$) with 
respect to $g_{(0)ij}$  using equations of motion.
When substituting the expression (\ref{vii}) 
into the classical action, the action diverges in general 
since the action contains the infinite volume integration 
on $M_{d+1}$. 
We regularize the infrared divergence by introducing a 
cutoff parameter $\epsilon$:
\be
\label{viii}
\int d^{d+1}x\rightarrow \int d^dx\int_\epsilon d\rho \ ,\ \ 
\int_{M_d} d^d x\Bigl(\cdots\Bigr)\rightarrow 
\int d^d x\left.\Bigl(\cdots\Bigr)\right|_{\rho=\epsilon}\ .
\ee
Then the action can be expanded as  power serie of 
$\epsilon$:
\bea
\label{viiia}
S&=&S_0(g_{(0)ij})\epsilon^{-{d \over 2}}
+ S_1(g_{(0)ij}, g_{(1)ij})\epsilon^{-{d \over 2}-1} \nn
&& + \cdots + S_{\rm ln} \ln \epsilon 
+ S_{d \over 2} + {\cal O}(\epsilon^{1 \over 2}) \ .
\eea
The term $S_{\rm ln}$ proportional to $\ln\epsilon$ 
appears when $d=$even. 
In (\ref{viiia}), the terms proportional to the inverse 
power of $\epsilon$ in the regularized action are invariant 
under the scale transformation 
\be
\label{viiib}
\delta g_{(0)\mu\nu}=2\delta\sigma g_{(0)\mu\nu}\ ,\ \  
\delta\epsilon=2\delta\sigma\epsilon \ . 
\ee
The invariance comes from the redanduncy (\ref{W7}). 
The subtraction of these terms proportional to the 
inverse power of $\epsilon$ does not break the invariance.
When $d$ is even, however, there appears the term 
$S_{\rm ln}$ proportional to $\ln\epsilon$.
The subtraction of the term $S_{\rm ln}$ breaks 
the invariance under the transformation(\ref{viiib}). 
The reason is that the variation of the $\ln\epsilon$ term 
under the scale transformation (\ref{viiib}) is finite 
when $\epsilon\rightarrow 0$ since 
$\ln\epsilon \rightarrow \ln\epsilon + \ln(2\delta\sigma)$. 
Therefore the variation should be canceled by the variation 
of the finite term $S_{d \over 2}$ (which does not depend 
on $\epsilon$) 
\be
\label{vari}
\delta S_{d \over 2}=-\ln (2\sigma)S_{\rm ln}
\ee
since the original total action (\ref{vi}) is invariant 
under the scale transformation. 
Since the action $S_{d \over 2}$ can be regarded as 
the action renormalized by the subtraction of the terms 
which diverge when $\epsilon\rightarrow 0$, 
the $\ln\epsilon$ term $S_{\rm ln}$ gives the conformal 
anomaly $T$ of the renormalized theory on the boundary 
$M_d$: 
\be
\label{xi}
S_{\rm ln}=-{1 \over 2}
\int d^2x \sqrt{-g_{(0)}}T \ .
\ee
 With above procedure, for the general action
\bea
\label{vi}
S&=&\int d^{d+1} x \sqrt{-\hat G}\left\{a \hat R^2 
+ b \hat R_{\mu\nu}\hat R^{\mu\nu}
+ c \hat R_{\mu\nu\rho\sigma}\hat R^{\mu\nu\rho\sigma}
+ {1 \over \kappa^2} \hat R - \Lambda \right\} \nn
&& + S_B\ ,
\eea
we find the following anomaly for $d=4$:
\be
\label{xvaa}
T=\left(-{l^3 \over 8\kappa^2}+5al+bl\right)(G-F) 
+ {cl \over 2}(G+F)\ .
\ee
Here we introduced the length parameter $l$ by 
\be
\label{ll}
l^2=-{{12 \over \kappa^2}\pm \sqrt{
{144 \over \kappa^4}-4d(d-3)\left\{(20a + 4b + 2c\right\}
\Lambda} \over 2\Lambda}\ .
\ee
The sign in front of the root in the above equation 
may be chosen to be positive which 
corresponds to the Einstein gravity ($a=b=c=0$).
We also used the Gauss-Bonnet 
invariant $G$ and the square of the 4d Weyl tensor $F$, 
which are given by
\bea
\label{GF} 
G&=&R_{(0)}^2 -4 R_{(0)ij}R_{(0)}^{ij} 
+ R_{(0)ijkl}R_{(0)}^{ijkl} \nn
F&=&{1 \over 3}R_{(0)}^2 -2 R_{(0)ij}R_{(0)}^{ij} 
+ R_{(0)ijkl}R_{(0)}^{ijkl} \ .
\eea
Especially for the action (\ref{W4}), we find 
($a={1 \over 6}\tilde c$, $b=-{4 \over 3}\tilde c$ and 
$c=\tilde c$)
\be
\label{W8}
T=\left(-{l^3 \over 8\kappa^2}- {\tilde c l \over 2}\right)(G-F) 
+ {\tilde cl \over 2}(G+F)\ .
\ee
We should note that $l$ is given by $l^2 = 
- {12 \over \kappa^2\Lambda}$ from (\ref{ll}), especially in case 
of (\ref{prmtr}),
\be
\label{spl}
l^2 = - {12 \over \kappa^2\Lambda}=1\ .
\ee

For the ${\cal N}=2$ theory with the gauge group $Sp(N)$, 
the usual UV Weyl anomaly is given by
\bea
\label{W11}
T&=&{1 \over 24\cdot 16\pi^2}\Bigl[ 
-\left(8N^2 + 6N - {1 \over 3}\right) R^2 \nn
&& + (24N^2+12N) R_{ij}R^{ij} + (6N-1)R_{ijkl}R^{ijkl}\Bigr] \\ 
&=& {1 \over 24\cdot 16\pi^2}\left\{ \left(-12N^2 -15 N 
+ {3 \over 2}\right) (G-F) 
+ \left(3N-{1 \over 2}\right)(G+F) \right\}\ .\nonumber 
\eea
Comparing (\ref{W8}) and (\ref{W11}), we find that the anomaly 
in (\ref{W8}) can be reproduced if
\be
\label{W12}
\tilde c l= {6N-1 \over 24\cdot 16\pi^2}\ ,\quad 
{l^3 \over \kappa^2}={12N^2 + 12 N -1 \over 3\cdot 16\pi^2}\ .
\ee
The second equation in (\ref{W12}) is, however, not compatible 
with (\ref{prmtr}), where ${l^3 \over \kappa^2}={N^2 \over 4\pi^2}$ 
since $l=1$ from (\ref{spl}). This might suggest that some 
sub-leading corrections to ${1 \over \kappa^2}$ and/or $\Lambda$ 
would be necessary. 
In other words, the bulk action (\ref{W4}) may be dual to 
${\cal N}=2$ $Sp(N)$ theory subject that some sub-leading 
corrections are induced to ${1 \over \kappa^2}$ (in minimal 
case) and to $\tilde c$. (One example of such sort is given 
by eqs.(23) with $l=1$ and cosmological constant equal to 
$-12$ of inverse gravitational constant.)  
The mechanism to produce such corrections is not yet clear. One 
possibility is that they come from field redefinitions.

In \cite{NO2}, the free energy of the corresponding ${\cal N}=2$ 
theory has been calculated from the Weyl gravity side (corresponding 
AdS black hole)
\be
\label{ff0}
F=- V_3 \left( - {\Lambda \over 12}\right)^{-3}
{\left(\pi T\right)^4 \over \kappa^8}\left(
1-{18\tilde c\Lambda \kappa^4 \over 12}\right)\ .
\ee
By substituting (\ref{W12}) into (\ref{ff0}) and fixing $l$ 
to be unity, we obtain
\be
\label{ff01}
F=-{\pi^2 V_3 N^2 T^4 \over 4}\left(1+{17 \over 8N} 
+ {\cal O}\left(N^{-2}\right)\right)\ .
\ee
On the other hand, from the field theoretical viewpoint (UV calculation 
of free energy in corresponding quantum field theory),
 we 
obtain 
\be
\label{ff1}
F=-{\pi^2 V_3 N^2 T^4 \over 3}\left(1+{2 \over N}-{1 \over 4N^2}
\right)\ .
\ee
The coefficient ${17 \over 8}$ of next-to-leading term in 
(\ref{ff01}) is approximately $2$ in the corresponding one in 
(\ref{ff1}) besides the usual overall factor ${4 \over 3}$.  Such close 
correspondence indicates again that Weyl gravity action under discussion is 
indeed supergravity dual of super YM theory with two supersymmetries.

In case of ${\cal N}=4$ theory, 
the term proportional to $(G+F)$ does not appear in the Weyl anomaly. 
This requires $\tilde c=0$. Therefore the anomaly in ${\cal N}=4$ 
theory cannot be reproduced only 
by the squared Weyl tensor term.

More generally we can consider the ${\cal N}=2$ $Sp(N)$ theory with 
$n_V=2N^2 + N$ vectormultiplets and $n$ sets of ($n_H
=(2N^2 + 7N -1)n$ hypermultiplets 
\bea
\label{W13}
T&=&{1 \over 24\cdot 16\pi^2}\Bigl[ 
-\left({11n_V \over 3} + {n_H \over 3}\right) R^2 \nn
&& + 12n_H R_{ij}R^{ij} + (n_H - n_V)R_{ijkl}R^{ijkl}\Bigr] \\ 
&=& {1 \over 24\cdot 16\pi^2}\left\{ \left(-{9n_V \over 2} 
-{3n_H \over 2} \right) (G-F) 
+ \left(-{n_V \over 2}+{n_H \over 2}\right)(G+F) \right\} \nn
&=& {1 \over 24\cdot 16\pi^2}\left\{ \left(-{9(2N^2 + N) \over 2} 
-{3(2N^2 + 7N -1)n \over 2} \right) (G-F) \right. \nn
&& \left. + \left(-{2N^2 + N \over 2}+{(2N^2 + 7N -1)n \over 2}
\right)(G+F) \right\} \ .\nonumber 
\eea
By comparing (\ref{W8}) and (\ref{W13}) again, we find that the 
anomaly in (\ref{W8}) can be reproduced if
\bea
\label{W14}
\tilde c l&=& {-n_V + n_H \over 24\cdot 16\pi^2}
= {-2N^2 - N + (2N^2 + 7N -1)n \over 24\cdot 16\pi^2} \nn
{l^3 \over \kappa^2}&=&{5n_V + n_H \over 3\cdot 16\pi^2}
={5(2N^2 + N) + (2N^2 + 7N -1)n \over 3\cdot 16\pi^2}\ .
\eea
We again come to the same conclusion: bulk action (\ref{W4}) may be 
dual to ${\cal N}=2$ SCFT subject to existance of sub-leading 
corrections to ${1 \over \kappa^2}$ and/or $\Lambda$. Another 
alternative could be the presence of some extra terms (fields) in 
action (\ref{W4}). That requires very careful investigation of 
string effective action.

Acknoweledgements. We thank A.Tseytlin for useful discussions.

\end{document}